\documentclass[12pt]{iopart}
\usepackage{graphicx}

\begin{document}

\title{Multiple-electron losses of highly charged ions
colliding with neutral atoms}
\author{M S Litsarev$^1$, V P Shevelko$^2$}
\address{$^1$Department of Physics and Astronomy, Uppsala University,
Box 516, 75120 Uppsala, Sweden}
\address{$^2$P.N. Lebedev Physical Institute, Leninskii prospect 53, 119991 Moscow, Russia}
\eads{\mailto{$^1$mikhail.litcarev@physics.uu.se},
\mailto{$^2$shev@sci.lebedev.ru}}

\begin{abstract}
We present calculations of the total and $m$-fold electron-loss cross sections
using the DEPOSIT code for highly charged $U^{q+}$ ions ($q=10,31,33$)
colliding with $Ne$ and $Ar$ targets at projectile energies $E=1.4$
and $3.5$~MeV/u.
Typical examples of the deposited energy $T(b)$ and $m$-fold ionization 
probabilities $P_m(b)$ used for the cross-section calculations
as a function of the impact parameter $b$ are given.
Calculated $m$-fold electron-loss cross sections are in a
good agreement with available experimental data.
Although the projectile charge is rather high,
a contribution of multiple-electron loss cross sections to the
total electron-loss cross sections is high: about 65~\% 
for the cases mentioned.

\end{abstract}


\submitto{\PS}

\maketitle

\section{Introduction}

Multi-electron losses (EL) of projectile positive and negative ions 
colliding with neutral atoms and molecules
play a key role in many problems of
plasma physics and physics of accelerators, especially 
at low and intermediate collision energies. 
A contribution of multi-electron processes to 
the total electron-loss cross sections can
be substantial and reach up to more than 50~\% even 
for highly charged projectile ions.
The total and $m$-fold EL cross sections can be 
calculated in the classical approximation using
the Classical Trajectory Monte Carlo approach~\cite{Olson2,
OlsonCTMC,OlsonBook} or the energy-deposition 
model~\cite{Paper1,nimb09,Paper2,Paper3} in the impact parameter 
representation providing an overall agreement with experimental data
within a factor of two.

In the present work the DEPOSIT code described in Ref.~\cite{cpc2012}
is used to calculate the total and $m$-fold EL cross sections for 
$U^{q+}$ ions, ($q=10,31, 33)$ colliding with $Ne$ and $Ar$
targets at energies $E=1.4$ and $3.5$~MeV/u.
The results are in a good agreement with available experimental 
data~\cite{DuBoispra2004,Perumal2005}.

\section{Theoretical approach}

The DEPOSIT code is based on the semi-classical energy deposition model
introduced by N.~Bohr~\cite{bohr}
and developed further in Ref.~\cite{Cocke1} for ionization
of \textit{atoms} by ions. 
Here, this model is used for ionization of \textit{ions}
by neutral atoms at energies about a few MeV/u, where
a contribution of multiple-electron loss processes to the total loss
cross-section is very large

In the energy deposition model, the problem of determination
the $m$-fold ionization probabilities $P_{m}(b)$ 
is reduced to calculation of a 3D integral over the ion-coordinate space 
using projectile electron density $\rho_{\gamma}(r)$ and 
the energy gain $\Delta E_{\gamma}(p)$ obtained
in the field $U(R)$ of the target atom
\begin{equation}
\label{Tb3DIntegral}
T(b)=\sum_{\gamma} \int \rho_{\gamma}(r)
\,\Delta E_{\gamma}(p)\, d^{3} \mathbf{r}.
\end{equation}
Here, index $\gamma$ labels $nl$-shell of the ion,
where $n$ is the principal quantum number and 
$l$ is the orbital quantum number. The summation is over all occupied shells
of the projectile ion. The energy gain $\Delta E_{\gamma}$ is
determined depending on the relation between the ion velocity $v$
and orbital velocity $u_{\gamma}=\sqrt{2I_{\gamma}}$,
$I_{\gamma}$ is the binding energy of the projectile $\gamma$-shell.

Electron density $\rho_{\gamma}(r)$ in the present work is calculated using 
nodeless Slater wave functions, and
atomic field at a distance $R$ from its nucleus is given by 
sum of three Yukawa potentials
with five fitting parameters defined from
the Dirack-Hartree-Fock-Slater calculations~\cite{DHFS_fitt}.

The total EL cross-section is obtained directly from the equation
\begin{equation}
\label{sigmatotEq}
\sigma_{tot}(v)=\sum_{m=1}^{N} \sigma_m = \pi b_{\max}^{2},
\end{equation}
where parameter $b_{\max}$ is (numerically) found from the equation
\begin{equation}
\label{bmaxEquation}
T(b_{\max})=I_{1},
\end{equation}
using the first ionization potential $I_1$ of the projectile ion
and expression~(\ref{Tb3DIntegral}) for the energy $T(b)$.

The $m$-fold cross-section is calculated in the following way
\begin{equation}
\label{sigmam}
\sigma_m(v)=2\pi \int^{\infty}_{0} P_m(b) b\, db,
\end{equation}
where the $m$-fold ionization probabilities $P_{m}(b)$
are taken from the statistical Russek-Meli model~\cite{Russek2}.
In this model the $P_m(b)$ probabilities are 
presented analyticaly and can be found by
using of the deposited energy $T(b)$
and the ionization potentials~$I_{k}$ of the projectile.

\section{Results and discussions}

\begin{figure*}
\includegraphics[width=10cm]{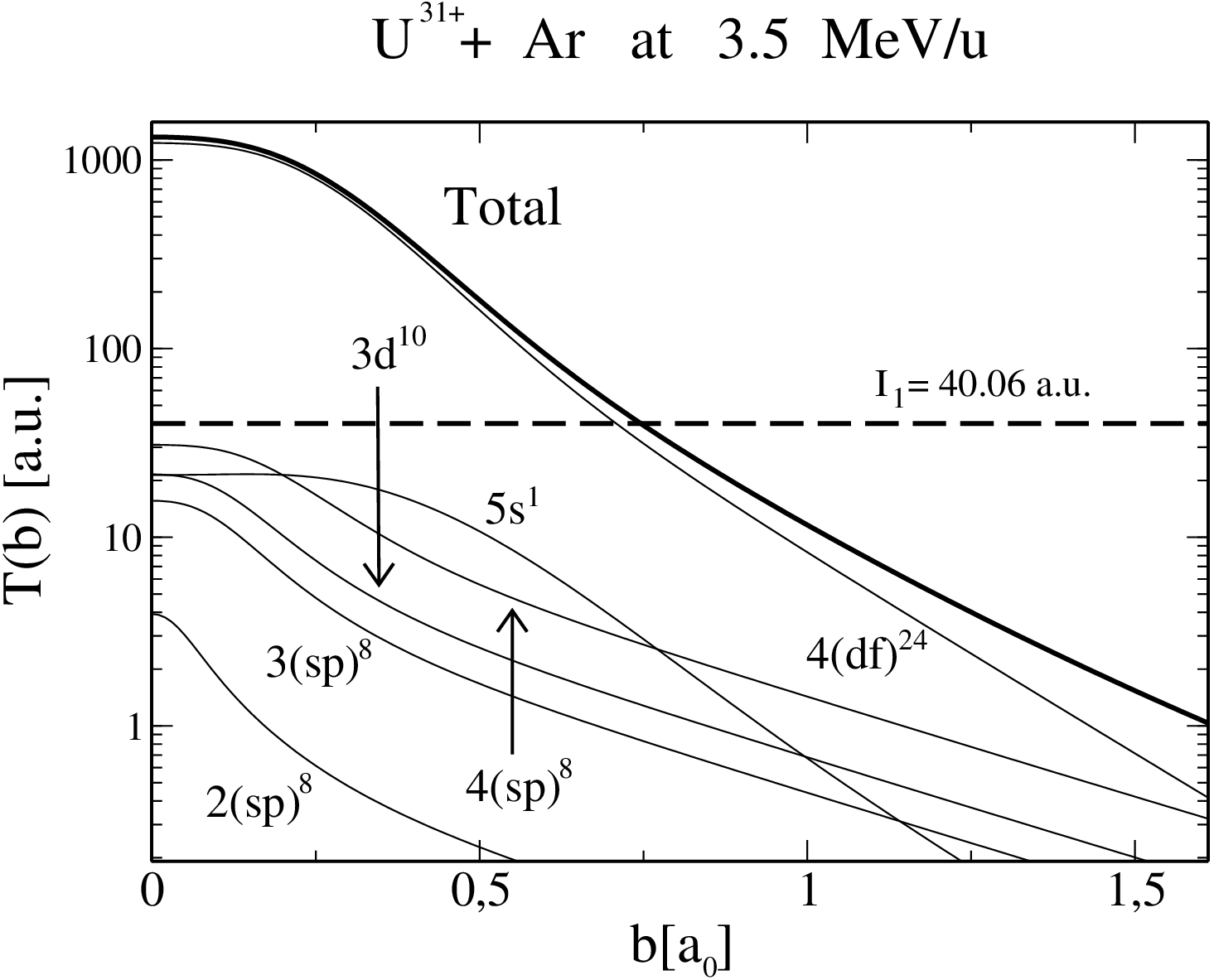}
\caption{ 
\label{fig1Tb}
Calculated energy $T(b)$ deposited to uranium ions 
by $Ar$ atoms as a function of the impact parameter $b$,
the DEPOSIT code. A contribution of the deposited energy 
to different subshells of uranium ions are shown.
The horizontal line corresponds to the first ionization potential
of $U^{31+}$: $I_1=40$ a.u. (1090 eV).
The notation for the Slater shells, for example,
$2(sp)^8$ means the electron configuration of the $2s^22p^6$ shell.
}
\end{figure*}

\begin{figure*}
\includegraphics[width=10cm]{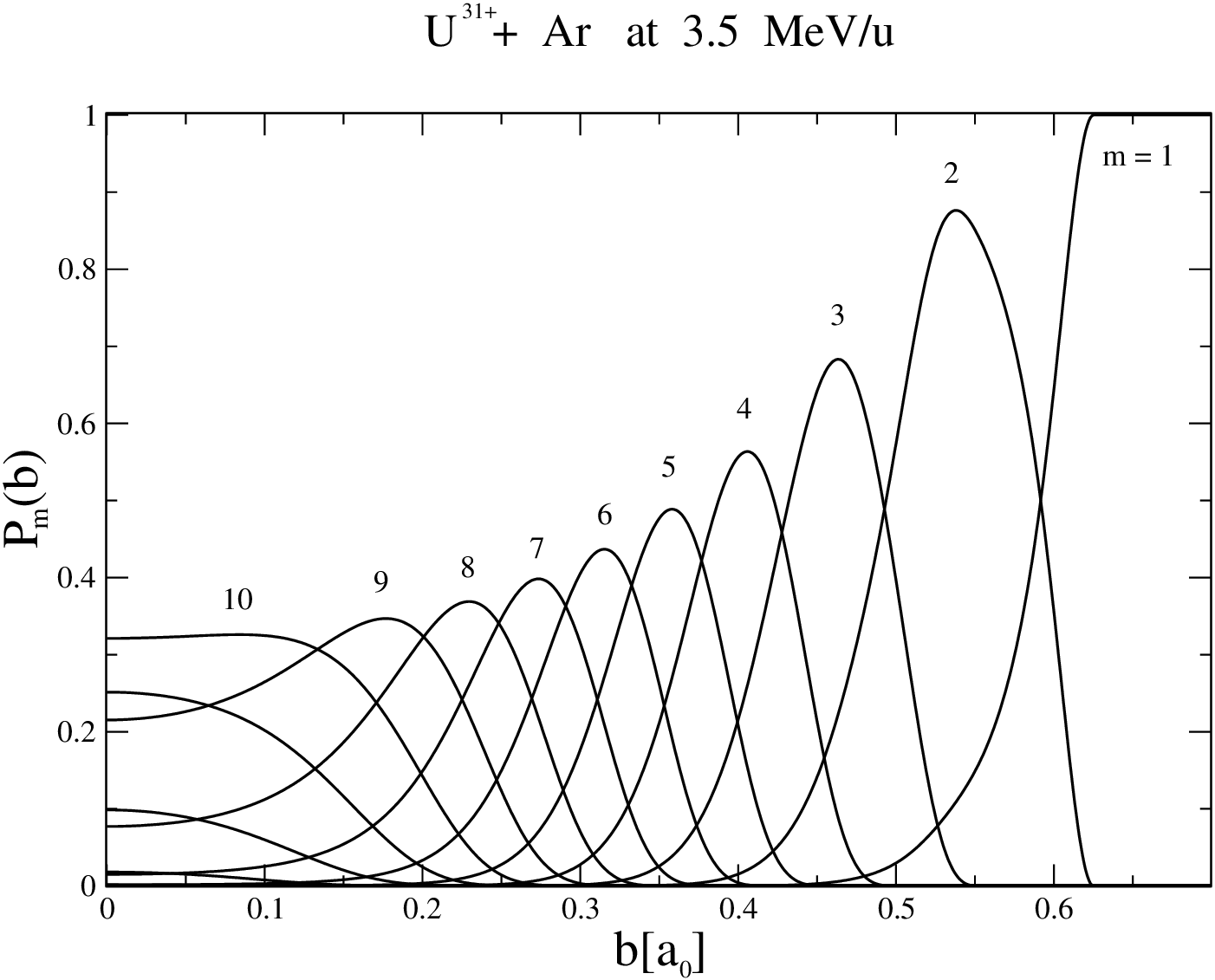}
\caption{
\label{fig2pm}
Calculated probabilities $P_m(b)$ for the $m$-fold ionization of $U^{31+}$ ions 
colliding with $Ar$ atoms at energy $E=1.4$ MeV/u as a function of impact 
parameter~$b$. In every $b$-point the condition $\sum_{m}P_m(b)=1$
is satisfied.
}
\end{figure*}

\begin{figure*}
\includegraphics[width=10cm]{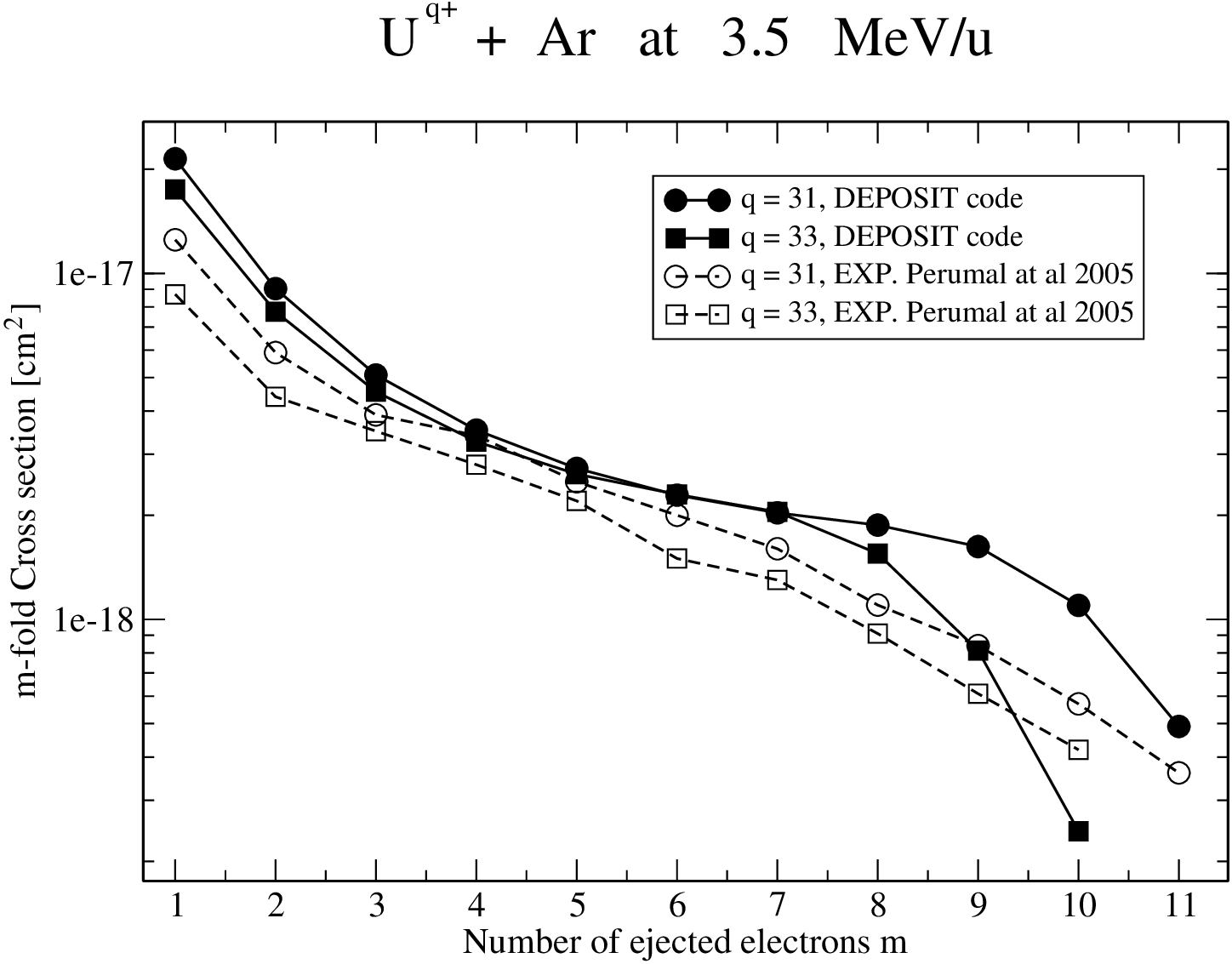}
\caption{ 
\label{fig3uqar}
Multiple-electron loss in $U^{q+} + Ar$ collision at energy $E = 1.4$~MeV/u.
The $m$-fold electron-loss cross sections as a function of ejected electrons $m$.
Experiment~-- Ref.~\cite{Perumal2005}, theory -- the DEPOSIT code.
The binding energies for  $U^{q+}$ ions are taken
from Ref.~\cite{Rasid}.
}
\end{figure*}

\begin{figure*}
\includegraphics[width=10cm]{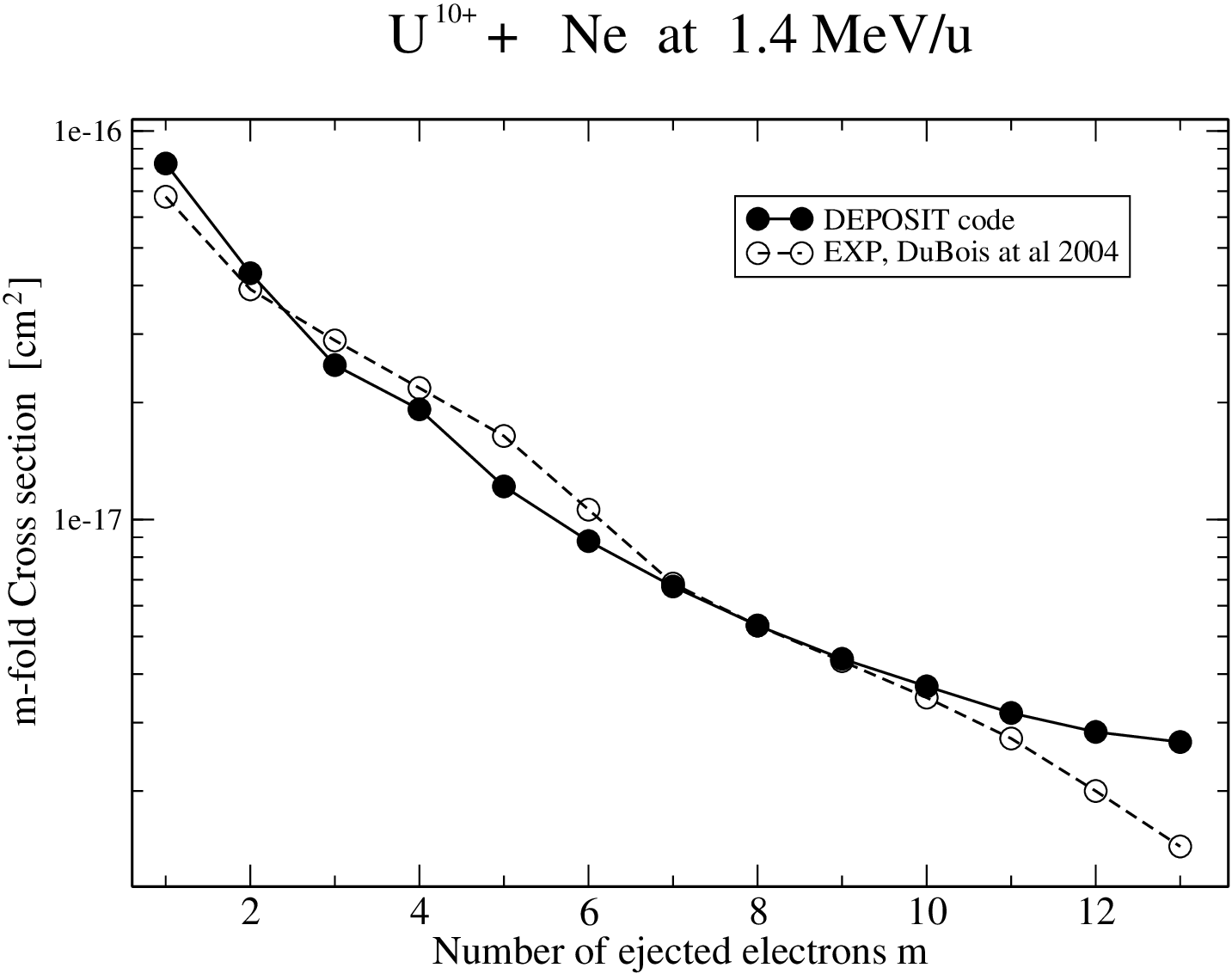}
\caption{ 
\label{fig4u28xe}
The $m$-fold electron-loss cross sections of
$U^{10+}$ ions colliding with $Ne$ atoms
at the collision energy $E = 1.4$~MeV/u.
Experiment~-- Ref.~\cite{DuBoispra2004}, theory -- the DEPOSIT code.
The binding energies for  $U^{10+}$ ions are taken
from Ref.~\cite{Rasid}.
}
\end{figure*}

The calculated deposited energies $T(b)$ and ionization probabilities
$P_{m}(b)$ are presented for $U^{31+}$ ions colliding 
with $Ar$ target at the energy $E=3.5$~MeV/u in Fig.~\ref{fig1Tb}
and Fig.~\ref{fig2pm} correspondingly.
The calculated $m$-fold EL cross sections for uranium
ions $U^{q+}$ ($q=10,31,33$) colliding
with $Ne$ and $Ar$ atoms are given in
Fig.~\ref{fig3uqar} and Fig.~\ref{fig4u28xe} in
comparision with experimental data.

Fig.~\ref{fig1Tb} shows typical example of the $T(b)$
behaviour. 
The deposited energy for $U^{31+}$ ion is transferred
mainly to the $4d^{10}4f^{14}$ shell for all
valid values of the impact parameter~$b$.
The value $b_{\max}=0.76$~$a_0$ is obtained 
from the equation~(\ref{bmaxEquation}) for
the first ionization potential $I_1=40.06$~a.u.
of $U^{31+}$ ion. For this system 
the inner-shell electrons does not contribute much
to the ionization process.

Ionization probabilities calculated for
$U^{31+}+Ar$ system colliding at the energy $E=3.5$~MeV/u
are presented in Fig~\ref{fig2pm}.
The Figure shows that for a given collision energy the probability
of the $12$-fold and more ionization processes 
is too small. This result is in correspondence with the experimental
data~\cite{Perumal2005} available only for 
number of ejected electrons $m\le11$.

In the Fig~\ref{fig3uqar} and
Fig~\ref{fig4u28xe}
the calculated $m$-fold EL cross sections are shown
in comparison with experiment.
As one can see, the difference between the two data sets
(experimental and theoretical) is small.
The figures show that multi-electron losses contribute to the
total EL loss significantly.

Finally, we present calculated total EL cross sections 
for all the considered systems.
For $U^{31+}+Ar$ at $E=3.5$~MeV/u
$\sigma_{tot}=5.18\times 10^{-17}$~$cm^2$
(experimental value
$\sigma^{exp}_{tot}=3.48\times 10^{-17}$~$cm^2$),
for $U^{33+}+Ar$ at $E=3.5$~MeV/u
$\sigma_{tot}=4.27\times 10^{-17}$~$cm^2$
($\sigma^{exp}_{tot}=2.63\times 10^{-17}$~$cm^2$),
and for
$U^{10+}+Ne$ at $E=1.4$~MeV/u
$\sigma_{tot}=2.27\times 10^{-16}$~$cm^2$
($\sigma^{exp}_{tot}=2.09\times 10^{-16}$~$cm^2$).
The corresponding experimental data are
taken from the Refs.~\cite{DuBoispra2004,Perumal2005},
and are in a reasonable agreement with results of our
calculations.

\section{Conclusions}

As seen from the experimental data for $m$-fold EL cross sections, 
a contribution of muti-electron losses ($m > 1$) to the 
total EL cross sections is very large and in the cases considered 
here reaches about 65~\%.  Calculations performed by the DEPOSIT code 
well reproduce this result. It means that multi-electron loss processes 
should be taken into account in investigations of kinetic of highly charged ions,
first of all, in determination of the ion-charge fractions, mean charges of ions passing through the solid, gaseous or plasma targets, in estimations of the vacuum conditions in accelerator machines and many others.

\ack
The computations were performed on resources  provided
by the Swedish National Infrastructure for Computing (SNIC)
at Uppsala Multidisciplinary Center
for Advanced Computational Science (UPPMAX)
under the Project \textit{s00311-8}
and on resources provided by SNIC
at the National Supercomputer Center (NSC)
under the Project \textit{matter2}.

This work was partly supported by the grant 
\#\textit{11-02-00526a}
of the Russian Foundation for Basic Research.

\section*{References}

\begin{thebibliography}{99}
\bibitem{Olson2}
Olson R E \textit{et al} 2004 \textit{\jpb} \textbf{37} 4539.
\bibitem{OlsonCTMC}
Olson R E 1989 \textit{\PR}A \textbf{39} 5572.
\bibitem{OlsonBook}
Olson R E 1996 \textit{Atomic, Molecular, and Optical Physics Handbook}
(College Park) chapter 56.
\bibitem{Paper1}Shevelko V P \textit{at al} 2008
\textit{\jpb} \textbf{41} 115204.
\bibitem{nimb09}
Song M-Y \textit{at al} 2009 \textit{\NIM}B \textbf{267} 2369.
\bibitem{Paper2}
Shevelko V P \textit{at al} 2009 \textit{\jpb} \textbf{42} 065202.
\bibitem{Paper3}
Shevelko V P \textit{at al} 2010 \textit{\jpb} \textbf{43} 215202.
\bibitem{cpc2012} 
Litsarev M S 2012 (submitted) \textit{Comp. Phys. Com.},
\textit{arXiv:1205.5390}.
\bibitem{DuBoispra2004} 
DuBois R D \textit{at all} 2004 \textit{\PR}A \textbf{70} 032712.
\bibitem{Perumal2005} 
Perumal A N \textit{at al} 2005 \textit{\NIM}B \textbf{227} 251.
\bibitem{bohr}
Bohr N 1915 \textit{Phil. Mag.} {\bf 30} 581.
\bibitem{Cocke1}
Cocke C L 1979 \textit{\PR}A \textbf{20} 749.
\bibitem{DHFS_fitt}
Salvat F \textit{at al} 1987 \textit{\PR}A \textbf{36} 467.
\bibitem{Russek2}
Russek A and Meli J 1970 \textit{Physica} A {\bf 46} 222.
\bibitem{Rasid}
Rashid K \textit{at al} 1988 \textit{At. Data Nucl. Data Tables}
\textbf{40} 365.

\end{thebibliography}
\end{document}